\documentclass[fleqn]{article}
\usepackage{longtable}
\usepackage{graphicx}

\begin {document}
\begin{flushleft}
{\LARGE
{\bf ASSESSMENT OF ATOMIC DATA: PROBLEMS AND SOLUTIONS}
}\\

\vspace{1.5 cm}

{\bf {Kanti  M  ~AGGARWAL and Francis  P  ~KEENAN}}\\ 

\vspace*{1.0cm}

Astrophysics Research Centre, School of Mathematics and Physics, Queen's University Belfast, Belfast BT7 1NN, Northern Ireland, UK\\ 

\vspace*{0.5 cm} 

e-mail: K.Aggarwal@qub.ac.uk \\

\vspace*{1.50cm}

Received  21 November 2012\\
Accepted for publication  12  January 2013 \\
Published xx  Month 2013 \\
Online at  É..  \\

\vspace*{1.5cm}

PACS Ref: 32.70 Cs, 34.80 Dp, 95.30 Ky

\vspace*{1.0 cm}

\hrule

\vspace{0.5 cm}

\end{flushleft}

\clearpage


\begin{abstract}

For the reliable analysis and modelling of  astrophysical, laser-produced and fusion plasmas, atomic data are required for a number of parameters, including energy levels, radiative rates and electron impact excitation rates. Such data are desired for a range of elements (H to W) and their many ions. However, measurements of atomic data, mainly for radiative and excitation rates, are not feasible for many species and therefore calculations are needed. For some ions (such as of C, Fe and Kr) there are a variety of calculations available in the literature, but often they significantly differ from one another. Therefore, there is a great demand from the user community to have data `assessed'  for accuracy so that they can be confidently applied to the modelling of plasmas. In this paper we highlight the difficulties in assessing atomic data and offer some solutions for improving the accuracy of calculated results.

\end{abstract}

\clearpage

\section{INTRODUCTION}

Atomic data, including energy levels, radiative rates, and electron impact excitation rates, are required for the modelling of a variety of plasmas, such as  astrophysical, laser-produced and fusion. These data are also needed for determining plasma parameters, such as temperature, density and  chemical abundance. Since experimental values for most of the above atomic parameters are neither available nor can be easily measured, one has to depend on theory. For  ions of heavy elements (mainly $Z >$ 28)   there is a scarcity of atomic data, but for lighter ions often multiple calculations are available in the literature. However, significant differences  are frequently observed among different yet comparable calculations, for almost all required atomic parameters -- see, for example \cite{fe25}. For this reason there is a great demand from the user community to have  data `assessed'  for accuracy, so that they may be confidently applied.  However, there are many difficulties in assessing  atomic data. In this paper we discuss some of these difficulties and offer  possible solutions  for improving the accuracy of calculated data.

\section{ATOMIC PARAMETERS}

The atomic parameters most commonly required for the modelling of plasmas are the following. 

\subsection{Energy Levels} 
The transition energy E$_{ij}$ and wavelength $\lambda_{ij}$ are related by the following simple relationship:
\begin{equation}
E_{ij} = E_j - E_i = h\nu_{ij} = hc/\lambda_{ij}
\end{equation}
where $i$ and $j$ are the {\em lower} and {\em upper} levels of a transition, $h$ is Planck constant, $c$ is the velocity of light in vacuum, and $\nu_{ij}$ is the transition
frequency. Atomic physicists generally refer to transition energies, whereas astronomers are more comfortable using wavelength terminology. For many ions experimental values have been compiled and critically assessed by the National Institute of Standards and Technology (NIST). Their assessed data are regularly published in refereed journals and are also freely available on their website: {\tt http://www.nist.gov/pml/data/asd.cfm}. However, often their data are incomplete (i.e. there are many missing levels, see for example, He-like ions),  or the levels of a state have non-degenerate energies. In such circumstances there is no choice except to use theoretical energy values. 

\subsection{Radiative Rates}
Radiative rates (A- values), also known as Einstein A-coefficients or transition probabilities, are related to the absorption oscillator strengths (f- values) by the following
relationship:
\begin{equation}
f_{ij} = \frac{mc}{8{\pi}^2{e^2}}{\lambda^2_{ji}} \frac{{\omega}_j}{{\omega}_i} A_{ji}
 = 1.49 \times 10^{-16} \lambda^2_{ji} (\omega_j/\omega_i) A_{ji}
\end{equation}
where $m$ and $e$ are the electron mass and charge, respectively, and $\omega_i$ and $\omega_j$ are the statistical weights of the lower ($i$) and upper ($j$) levels, 
respectively. The f- values are dimensionless and A- values are in units of sec$^{-1}$.  Generally, A- values for electric dipole (E1) transitions are dominant but contributions from other type of transitions, namely   magnetic dipole (M1), electric quadrupole (E2) and magnetic quadrupole (M2), can be appreciable, and therefore are required in a complete plasma model. Measurements of A- values are sparse and therefore these are often determined theoretically. Unlike energy levels, the A- values are not  rigorously assessed by the NIST and the corresponding data available on their website are very limited.

\subsection{Lifetimes}
The lifetime $\tau$ for a level $j$ is defined as follows:
\begin{equation}  
{\tau}_j = \frac{1}{{\sum_{i}^{}} A_{ji}}.
\end{equation}
Since this is a measurable parameter, it provides a check on the accuracy of the calculations for A- values. However, measurements are confined only to a few levels of a limited set of ions. Theoretically, lifetime calculations include mainly the contribution of the E1 transitions, but those from other types of transitions may be significant - see, for example, Aggarwal and Keenan \cite{arxvii}--\cite{ovii}.

\subsection{Electron Impact Excitation Collision Strengths}

The collision strength ($\Omega$) is related to the better-known quantity collision {\em cross section} ($\sigma$) as follows:
\begin{equation} 
{\Omega}_{ij}(E) = {k^2_i}\omega_i\sigma_{ij}(\pi{a^2_0})
\end{equation}
where ${k^2_i}$ is the colliding energy of the electron, $\omega_i$ is the statistical weight of the lower ($i$)  level,  and ${a_0}$ is the Bohr radius. Since $\Omega$ is a {\em dimensionless} quantity, intercomparisons among various calculations become straightforward. As for A- values, measurements of $\sigma$ or $\Omega$ are very limited and therefore hardly provide any comparisons with theory. Furthermore, calculations and/or measurements for $\Omega$ at a few energies are not sufficient, as the thresholds energy region is often dominated by numerous closed-channel (Feshbach) resonances -- see Figs. 6--9 of \cite{fe25}. Therefore, values of $\Omega$ need to be calculated in a fine energy mesh in order to accurately account for their contribution. Furthermore, in a hot plasma electrons have a wide distribution of velocities, and therefore values of $\Omega$ are generally averaged over a {\em Maxwellian} distribution to determine the {\em effective} collision strengths as follows:
\begin{equation}
\Upsilon(T_e) = \int_{0}^{\infty} {\Omega}(E) {\rm exp}(-E_j/kT_e) d(E_j/{kT_e}),
\end{equation}
where $k$ is Boltzmann constant, T$_e$ is the electron temperature in K, and E$_j$ is the electron energy with respect to the final (excited) state. Once the value of $\Upsilon$ is known the corresponding results for the excitation q(i,j) and de-excitation q(j,i) rates can be easily obtained -- see Eqs. (9--10) of \cite{fe25}. Effective collision strengths  do not vary strongly with changing electron temperature, and therefore it is easier  to fit them to a polynomial function of T$_e$, as  in the {\sc chianti} database at  {\tt http://www.damtp.cam.ac.uk/user/astro/chianti/}. The contribution of resonances may enhance the values of $\Upsilon$ over those of the background values of collision strengths ($\Omega_B$), especially for the forbidden transitions, by up to an order of magnitude  (or even more) depending on the transition and/or the temperature.  Similarly, values of $\Omega$ need to be calculated over a wide energy range (above thresholds) in order to obtain convergence of the integral in Eq. (5).

\subsection{Line Intensity Ratio}

The intensity of an emission line can be expressed as:

\begin{equation}
I_{ji} = A_{ji}N_jN_{A,Z}N_{A}h{\nu}_{ji} \frac{n}{1+N_{He}} \frac{L}{4\pi} \hspace*{1cm} ergs \hspace*{0.15 cm} cm^{-2} s^{-1} sr^{-1} 
\end{equation}
where N$_j$ is the relative population of level j, N$_{A,Z}$ is the relative ionic abundance of ion with charge Z of element with atomic number A, N$_A$ is the relative (with
respect to hydrogen) chemical abundance, N$_{He}$ is the relative chemical abundance of helium, $n$ is the total number density of hydrogen and helium nuclei (in cm$^{-3 }$),
and L is the path length through the line emitting region. However, this equation applies to astrophysical plasmas whereas for laboratory and fusion plasmas some parameters, such as L and N$_{He}$ are not required. Nevertheles, calculations for the intensity of a single emission line requires many parameters, which subsequently add to the uncertainties. Therefore, the ratio of two lines $i \rightarrow j$ and $m \rightarrow n$ of an ion, i.e.

\begin{equation}
R = \frac{I({\lambda}_{ij})}{I({\lambda}_{mn})} = \frac{A_{ji}}{A_{nm}} \frac{{\lambda}_{mn}}{{\lambda}_{ij}} \frac{N_j}{N_n}
\end{equation}
eliminates many parameters as it then  depends only on the wavelengths, A- values, and populations of the upper levels for which the above noted (and some other parameters, such as ionisation, recombination, and photoexcitation cross sections) are required. This is the main reason that a set of lines is often used as plasma diagnostics. If the two lines of an ion have common upper levels (i.e. $j$ = $n$), then R normally depends only on the wavelengths and A- values, except e.g. in the presence of opacity. However, such lines are not very useful as diagnostics because they are independent of the density and/or temperature of the plasma. Those  which are useful  vary with either density or temperature.

\section{CODES FOR CALCULATIONS OF ATOMIC DATA} 

For calculations of the above atomic parameters a variety of structure and scattering codes are available. Some of the most commonly used atomic structure codes to generate energy levels and A- values are: {\em Configuration Interaction Version 3} (CIV3: \cite{civ3}),  {\em General-purpose Relativistic Atomic Structure Package} (GRASP: \cite{grasp0}), {\em SuperStructure} (SS: \cite{ss}), {\em AutoStructure} (AS: \cite{as}), {\em Multi-Configuration Hartree-Fock} (MCHF: \cite{mchf}), {\em Many-Body Perturbation Theory} (MBPT: \cite{mbpt}), {\em Hartree-Fock Relativistic} (HFR: \cite{hfr}), {\em Flexible Atomic Code} (FAC: \cite{fac}), and {\em Hebrew University Lawrence Livermore Atomic Code} (HULLAC: \cite{hullac}). Some of these codes have been published (such as CIV3, SS and GRASP), some are available on request (HULLAC), and some are on the web (FAC: {\tt {\verb+http://sprg.ssl.berkeley.edu/~mfgu/fac/+}}). Similarly, the most commonly used scattering methods are the $R$-matrix and {\em distorted-wave} (DW). The associated codes are the standard $R$-matrix (RM: \cite{rm}), which also incorporates the one-body relativistic operators, and the fully relativistic version, i.e. the {\em Dirac Atomic $R$-matrix Code} (DARC: {\tt http://web.am.qub.ac.uk/DARC/}). There are many versions of the DW method, but the most commonly used are those of University College London \cite{dw}, FAC and HULLAC. Furthermore, FAC and HULLAC are self-sufficient codes providing data for a range of atomic parameters, whereas the RM and DARC require  input wavefunctions from CIV3 (although SS and AS can also be used) and GRASP, respectively. Finally, it must be stressed that some codes are comparatively easier to use than others (particularly FAC), but with some practice it is not too difficult to obtain the required results from the other codes. However, the desired level of accuracy for the results is difficult to achieve. It is even more difficult to determine the accuracy, as we will discuss later. Finally, almost all codes are under constant development and therefore the latest (and the most suitable) version may differ from those published or made available earlier.

\section{CHOICES FOR A MODEL CALCULATION} 

For any model calculation a choice has to be made for a number of parameters some of which are listed below.

\begin{enumerate}

\item Number of states/levels
\item Configuration interaction (CI)
\item Inclusion/exclusion of pseudo states
\item Inclusion/exclusion of relativistic effects (RE) 
\item Energy and temperature ranges
\item Number of partial waves
\item Top-up
\item Inclusion/exclusion of resonances
\item Elimination of pseudo/spurious resonances
\item Inclusion/exclusion of radiation damping

\end{enumerate}

The inclusion/exclusion of the above parameters is generally crucial for the accuracy of the generated parameters. However, which to include/exclude and at what level/s depends on the requirements, such as: (i) where the data are to be applied, (ii) what level of accuracy is desired or acceptable, (iii) what codes are available to the worker/s and/or with what codes they are familiar with, and most importantly (iv) what computational resources and time-frame are available. Depending on these requirements a choice has to be made before beginning a calculation, particularly for the determination of $\Upsilon$. This is because some calculations  are so computationally challenging and time consuming that they cannot be easily repeated. For the same reason, i.e. to economise on computational resources, some of the above noted factors have to be compromised, which often leaves scope for improvements in subsequent work. However all the above parameters may not be important for a particular species. For example, CI is generally more important for lighter ions (around $Z \le$ 20) - see, for example, Aggarwal  \cite{c-like}, but may  sometimes be equally important for heavier ions - see, for example, Aggarwal et al \cite{ni-ions} for energy levels of Ni XIII to Ni XVI. In some cases, particularly for H-like ions, the inclusion of a large number of configurations is either not sufficient or not feasible, and therefore pseudo states need to be included \cite{bbs}  to achieve the desired accuracy in the derivation of energy levels and subsequent parameters. Often, extensive CI is included for the determination of energy levels and radiative rates, but restrictions are imposed for the further calculation of collisional data due to the limitation of computational resources. However, this approach leads to the generation of pseudo resonances \cite{oiii} which need to be smoothed over a wide energy range  to avoid the overestimation  of $\Upsilon$.

Another important contribution to the determination of energy levels is from the relativistic operators, namely mass correction, spin-orbit interaction, spin-other-orbit interaction, spin-spin interaction, Breit interaction, Darwin term, and quantum electrodynamics (QED) effects. Generally, the heavier the ion the more important is the contribution of these operators -- see for example \cite{kr35}. However, sometimes their significance is appreciable even for lighter ions, as shown by Aggarwal et al \cite{mg-like}. Particularly important is their contribution in splitting the fine-structure levels of a term - see, for example, Hamada et al \cite{hydro} for energy levels of H-like ions. Finally, for most of the lighter ions ($Z \le$ 20) inclusion of one-body relativistic operators, in a Breit-Pauli approximation, is sufficient as noted by \cite{caxix}, but for heavier ions fully relativistic calculations are preferable \cite{fe25},\cite{kr35}.

For collisional calculations the choices are crucial for the inclusion of: (i) the number of states/levels, (ii) the number of partial waves, and (iii) the energy
range up to which values of $\Omega$ are calculated. This is because electron-ion scattering, especially at low energies, can be considered as a two-step process, i.e. a temporary capture of the colliding electron by the target, followed by autoionization \cite{rm}: 

\begin{equation}
e^{-} + A_i \rightarrow (A^{-})^{\ast} \rightarrow e^{-} + A_j
\end{equation}
where the asterix indicates the (N+1) resonance states. It is the description of such processes which is explicitly included in $R$-matrix, but is (generally) ignored in the DW method. The contribution of resonances may dominate the determination of $\Upsilon$ at lower temperatures, particularly when they arise close to the thresholds - see, for example, Figs. 1-3 of Aggarwal \cite{nev} for transitions among the fine-structure levels of the 1s$^2$2s$^2$2p$^2$ ground configuration of Ne V. For these transitions, resonances have enhanced the values of $\Upsilon$ by over an order of magnitude at a temperature of 10$^4$ K. However, sometimes the contribution of resonances can be significant even at temperatures as high as 10$^6$ K, as demonstrated by Aggarwal and Keenan \cite{mo34} for transitions in Mo XXXIV and by Aggarwal et al \cite{gd37} for Gd XXXVII. This is particularly true when the energy difference between any two levels is very large. For example, the 2s$^2$2p$^5$ $^2$S$_{1/2}$ and 2s$^2$2p$^4$3s $^4$P$_{5/2}$ levels of Mo XXXIV are separated by over 155 Ryd (see Table 1 of \cite{mo34}), and the entire energy region is dominated by resonances, as shown in Fig. 1 of Aggarwal and Keenan \cite{mo34}.

Certain types of radiative rates from recombination resonances increase with ionization stage, while the autoionizing rates remain relatively constant \cite{dpz}. As a result, in many highly ionized species radiative rates from resonances can become comparable to the autoionizing rates, i.e. the dielectronic recombination competes with electron impact excitation, and hence can cause a significant reduction in the resonance contributions. This  effect is called radiation damping, and can be significant for some transitions, as discussed and demonstrated by \cite{dpz},\cite{icft} for He-like ions and by Ballance and Griffin \cite{w47} for W XXXXVII. Nevertheless, a majority of transitions are (generally) not affected by radiation damping, particularly at the high temperatures at which data are required for plasma modelling, as demonstrated by \cite{icft},\cite{damp}  and discussed by Aggarwal and Keenan \cite{fe25},\cite{kr35}. 

\section{ASSESSMENT OF ATOMIC DATA}

The assessment of atomic data is a tedious and never ending job, as newer and newer data keep appearing in the literature. However,  it requires expertise and experience to assess any data. Furthermore, it is not always easy to get the assessed data published, which inhibits the workers to undertake the job. Nevertheless, several institutions have undertaken the work of data assessment in the past. For example, in the 1970s, Los Alamos National Laboratory (USA) collected and assessed atomic data,  Queen's University Belfast (UK) did the similar work in the 1980s, and in the 1990s National Institute for Fusion Science (Japan) was quite active in storing, assessing and publishing  assessed data. At present, the National Institute of Standards and Technology (USA) is the only one active in storing, assessing and disseminating  data, but this is restricted to  energy levels and is mostly from measurements, although they do also compile some A- values. As there has always been a great demand from the user community to have assessed data, several databases have appeared in the past to fill the void, but only partially -- see section 8 below. However, in the recent past with the efforts of the IAEA, and keeping in view the future requirement of the fusion community (particularly ITER), the National Fusion Research Institute (South Korea) has voluntarily offered its services for data assessment. 

One of the major difficulties in assessing atomic data, particularly for radiative rates and collisional parameters ($\Omega$ and $\Upsilon$), is the lack of measurements, because these are not easy to perform.  Furthermore, even if  laboratory measurements are available for a few transitions and at a few energies for $\Omega$, these are not very helpful to assess $\Upsilon$ as data are required over a wide range of energies. Therefore, for most of the ions theoretical data are the only choice. Where a few calculations are already available, as  for a few Fe ions, it is comparatively  easier to assess the accuracy of the data. The major difficulty for assessment arises when there is a single calculation -- see section 7 below for details. Some of the calculations are so computationally demanding and time consuming that they require months (if not years) of work, even with the best resources  available -- see for example \cite{w47} for W XXXXVII. Since it is not feasible to repeat the calculations, if an error occurs in the basic input data, it cannot be easily corrected. However, in a majority of cases large errors can be avoided simply by making extensive comparisons. If no previous calculation is  available with which to compare then it is advisable to perform a parallel calculation with the {\sc fac} code, as it is freely available, easy to implement and quick to run.  The  alternative is to do some modelling with the help of databases and to compare the end results, such as line intensities or their ratios, with the observations. However, this approach is not always helpful as observations may be scarce,  taken at different times with separate instruments and cover different wavelength ranges, or lines may be blended. Nevertheless, we discuss below some of the errors commonly noticed in atomic data and offer the reasons for likely discrepancies between calculations.

\section{TYPES OF ERRORS}

Generally, large discrepancies between any two sets of data are due to three types of errors, namely (i) inherent error/s in the code, (ii) non (rigorous) assessment of data, and (iii) the (unjustified) approximations made in the calculations. The first type is difficult to detect but easier to correct. Often errors are noted with the repeated use of code/s for a variety of ions or following comparisons of results with other similar calculations, as noticed for He-like \cite{fe25},\cite{caxix},\cite{tixxi} and Li-like \cite{lia},\cite{lib} ions. This is the main reason that actively used codes are often under continuous development.  The second type of error appears mainly when author/s either make unrealistic  assumptions or assess the accuracy of their data based on expectation rather than rigorous tests and comparisons. As examples, see \cite{ni19a},\cite{ni19b} for Ni XIX and \cite{nixi} for Ni XI, which demonstrate the inadequate inclusion of partial waves and the energy ranges. Finally, in all calculations some approximations have to be made, as already discussed in section 3. This (generally) leaves scope for further improvement/s in the calculated data, and hence is a continuous process, but large discrepancies are rarely   noted.

Before performing calculations for an ion it is important to keep the application in mind, because different plasmas have different requirements. For example, many photoionized astrophysical plasmas (such as H II regions and planetary nebulae) have electron temperatures below 50,000 K, and hence the position of near-threshold resonances are very important because of their dominant contribution at low temperatures -- see Figs. 1-3 of Aggarwal \cite{nev} and a recent work by Palay et al \cite{o3}. On the other hand, collisionally dominated plasmas (such as solar and stellar coronae) have temperatures $\sim$10$^6$ K, which means that the position of near-threshold resonances may not be crucial, but calculations for $\Omega$ need to be performed over a large energy range in order to achieve convergence of the integral in Eq. (5). Similarly, a choice of a larger model of the ion (i.e. enlarging the number of levels included in a calculation) may significantly alter (improve) the subsequent calculations of $\Upsilon$, because of the inclusion of resonances arising from the additional levels -- see particularly the recent work \cite{fe17}--\cite{fe11} on Fe ions. Finally, laser-produced and fusion plasmas may require the calculations of $\Upsilon$ up to $\sim$10$^8$ K, depending on the ion. As stated earlier, the contribution of resonances is more appreciable in calculations of $\Upsilon$ at comparatively lower temperatures, but their contribution at higher temperatures can also be significant as noted in the cases of Kr XXXII \cite{kr32}, Mo XXXIV \cite{mo34} and Gd XXXVII \cite{gd37}. Ions of some elements, such as C, Si and Fe, have applications in a variety of plasmas. However, a calculation performed with one particular application in mind (e.g. for astrophysical plasmas) may not be directly applicable to another type of plasma. Doing this may lead to large errors in the analysis.

\section{DISCREPANCIES IN ATOMIC DATA} 

In principle, calculations performed for any atomic parameter by using any method/code should give similar results (if not the same), provided the model sizes (and other associated parameters) are comparable. Unfortunately however, that is often not the case. Here we discuss the kind of discrepancies frequently found for different parameters and their likely causes. 

\subsection{Energy Levels}

As an example, energy levels of the 2s2p$^6$3$\ell$ configurations of Ni XIX from the CIV3 code are lower by up to 1.5 Ryd than those calculated by the GRASP and FAC codes, or experimentally compiled by NIST ({\tt http://www.nist.gov/pml/data/asd.cfm}) -- see Table 1 of Aggarwal and Keenan \cite{ni19c}. The main reason for such a large discrepancy is that Hibbert et al \cite{ni19d} focused their attention only on the lowest 27 levels of Ni XIX, but reported results for up to 37 levels. On the other hand, energy levels for Ni XVII from the CIV3 calculations \cite{ni17a}  differ by over 2 Ryd from other theoretical and experimental results,  as confirmed by our calculations not only from the GRASP and FAC codes, but also from the CIV3 -- see Table 5 of Aggarwal et al \cite{fe15}.  In this case  the differences in energy levels are not understandable and there must have been some error in the calculations of \cite{ni17a}. Similarly, energies of  \cite{icft} from the AS code for levels of He-like ions are higher by up to 2 Ryd \cite{kr35}, depending on the ion, because two-body relativistic operators were not included in the code. Thus there can be several reason/s for the discrepancy in the reported energy levels.

Sometimes there is a problem in the assignment of the orderings of energy levels, as seen in Table 1 of Gupta et al \cite{al-like} for Al-like ions. The $^2$D$^o_{3/2,5/2}$ energy levels of the 3p$^3$ and 3s3p($^3$P$^o$)3d configurations are interchanged in the MBPT calculations of Safronova et al \cite{saf} for a series of ions as confirmed by the experimental results as well as our calculations from three independent codes, namely CIV3, FAC and GRASP. The most likely reason for the interchange is that Safronova et al performed calculations for a large number of ions but investigated the orderings for only a few (lighter) ions. In many instances the orderings of levels are not the same for all ions in a series, and hence the discrepancy. Generally, listings at the NIST website are helpful in assigning the level orderings, but occasionally there are differences with the theory as noted in the cases of Kr XXXI and Kr XXXII \cite{kr-ions}. 

In all calculations, the designation of an energy level is (mainly) determined on the basis of the strength of its corresponding eigenvector. However, often and particularly for those ions whose levels extensively mix with different configurations, such as Fe XVI \cite{fe16a}, it is not always possible to assign a unique (unambiguous) designation for a level, because a single eigenvector may dominate for several levels. In such cases, the best one can state is that a particular level has such a $J$ value of parity even or odd, but the corresponding configuration from which it comes remains ambiguous, and subject to the interpretation of the individual author/s and/or user/s. 

\subsection{Radiative Rates}

If the same level of CI (and relativistic operators) are included in a calculation, the f- or A- values should agree within $\sim$20\%, irrespective of the method or the computer code employed, and this is true particularly for  strong (f $\ge$ 0.01) transitions. For weaker transitions, the discrepancies among different calculations can be large, often by an order of magnitude or even more depending on the transition. This is because weaker transitions are more sensitive to different levels of CI and/or their energies ($\Delta E_{ij}$), and for the same reason their length and velocity forms also often differ quite significantly. On the other hand, the strong transitions are generally more stable in magnitude, and their length and velocity forms also agree closely, which gives an indication of the accuracy of a calculation. However, discrepancies for strong transitions can also be large among different calculations as seen in Table 5 of Aggarwal and Keenan \cite{ni19c} for Ni XIX. The differences in f- values obtained from the GRASP and FAC codes are up to 50\% with those from  CIV3 \cite{ni19d}, particularly for  transitions involving levels 28 and higher, due to the corresponding differences in energy levels as discussed above.  Similarly, f- values from the CIV3 code \cite{fe9b}  differ from the GRASP and FAC results, by up to an order of magnitude, for many transitions of Fe IX as may be seen in Table 5 of Aggarwal et al \cite{fe9a}. These differences arise in spite of the fact that the energy levels of \cite{fe9b} are in close agreement with the experimental values, because of the `adjustment' of the Hamiltonian (known as `fine-tuning'). It is difficult to fully explain such large differences in A- values for so many transitions, but one can speculate on the reason/s. In most calculations from the CIV3 code, after a preliminary survey all those levels/configurations whose eigenvectors (mixing coefficients) are very small (say $<$ 0.01) are removed from the final calculations in order to economise on computational effort, and this process affects the weaker transitions more than the stronger ones, because of the additive or destructive effect of the components. The choice of an appropriate  cut-off level below which the levels/configurations are removed depends on the authors/s and/or on the size of the calculations, and we believe this is the main reason for the large differences in f- values discussed above for transitions in Fe IX. Therefore, from this (and many other similar) examples we may conclude that the process of fine-tuning may make theoretical energy levels more accurate in magnitude, but not necessarily the subsequent calculations of radiative and collision rates. Similarly, the orderings of the levels may remain uncorrected in the absence of  experimental energies for a larger number of levels, if not all.

Differences in A- values for several (particularly weaker) transitions of Li-like ions are up to three orders of magnitude, as demonstrated by Aggarwal and Keenan \cite{lia}. These types of large discrepancies are difficult to detect without performing independent calculations using different code/s. While the application of inaccurate data may affect the modelling of plasmas, it may not always be possible to perform calculations with different codes. However, caution may be exercised by performing several tests with the same code.

\subsection{Collision Strengths and Effective Collision Strengths}

Adopting the same level of complexity, different scattering methods and/or codes should provide comparable values of $\Omega$, for a majority of transitions. However, differences among several comparable calculations are often abnormally striking as we discuss here with some examples. As shown in Fig. 4 of Aggarwal and Keenan \cite{fe16b}, values of $\Omega$ from the DW calculations \cite{fe16c} are significantly lower than those from {\sc darc} for several forbidden transitions of Fe XVI, and the discrepancy between the two calculations increases with increasing energy. This is because Cornille et al \cite{fe16c} included only a limited number of partial waves in their calculations with the assumption this would  be  sufficient for the convergence of $\Omega$, which is not the case as may be seen from Figs. 1--3 of  Aggarwal and Keenan \cite{fe16b}. Eissner et al \cite{fe16d}  corrected this limitation in their calculations from the $R$-matrix code, but their subsequent results of $\Upsilon$ are still underestimated for several transitions  over a wide range of temperature as seen in Fig. 12 of Aggarwal and Keenan \cite{fe16b}. This is because their calculations involved only the lowest  21 levels of Fe XVI, while the larger calculation of Aggarwal and Keenan  \cite{fe16b} with 39 levels also included resonances arising from the higher levels. However, even the results of  Aggarwal and Keenan may be further improved in a similar way by the inclusion of yet higher levels of the $n \ge$ 6 configurations of Fe XVI. Improvements over a previous calculation is a continuous process, particularly  with the increasing availability of computational resources.

Fe XVI is a moderately heavy ion for which the contribution of relativistic effects is important, but not dominant, in the calculation of atomic parameters. However, the semi-relativistic $R$-matrix calculations of Bautista \cite{fe16e} for $\Upsilon$ differ from the fully relativistic results from {\sc darc} by up to an order of magnitude, for several transitions, as noted in Table 2 of Aggarwal and Keenan \cite{fe16f}. This is because of the sudden shift in the background values of $\Omega$ ($\Omega_B$) by the inclusion of relativistic effects, as shown in Fig. 3 of Bautista \cite{fe16e}. The shifts in $\Omega_B$ are upwards, downwards, and random, and are not realistic, but happened because of an error in the adopted version of the code which the author was not aware of at the time of the calculations. Most of the atomic codes currently in use are (generally) always under continuous development, as already stated and is illustrated by the several versions of the {\sc fac} code available on the web. An error in a code is often difficult to spot and harder to speculate in advance, and can only be noted and corrected after its prolonged use for a variety of calculations, and mainly by comparisons with other similar but independent calculations.

Generally, differences of up to a factor of two or so in values of $\Omega$ between any two calculations, at a few energies, do not considerably affect the subsequent determination of $\Upsilon$. Nevertheless, discrepancies in  $\Upsilon$ are sometimes up to three orders of magnitude, as recently demonstrated by Aggarwal and Keenan \cite{fe25} for transitions of several He-like and Li-like ions \cite{lia},\cite{lib}. Particularly affected are those transitions which belong to the degenerate levels of a state (often referred to as `elastic'  transitions because of their very small energy differences). Furthermore, these differences in $\Upsilon$ are not confined to the lower or higher values of temperature, but persist over the entire range as can be seen in Figs. 11 and 12 of Aggarwal and Keenan \cite{mgxi}. Such large discrepancies in  $\Upsilon$ reported by Whiteford et al \cite{icft} happened, yet again, due an error in the adopted version of the $R$-matrix code, and has now been corrected \cite{lb}. However, without the availability of independent calculations with the {\sc darc} code, it was not possible to know the errors in the reported data for several He-like \cite{icft} and Li-like \cite{ar16} ions.

Calculations of $\Omega$ (and subsequently $\Upsilon$) for allowed transitions among degenerate levels of a state, such as 2s$_{1/2}$--2p$_{1/2,3/2}$ in H-like ions, are  difficult to perform, because of their very slow convergence \cite{hydro} with partial waves. Additionally, a slight variation in their $\Delta E_{ij}$ may easily lead to an under/overestimation in  $\Omega$ and subsequently of $\Upsilon$, as discussed and demonstrated in detail by Hamada et al \cite{hydro}. 

Even without an error in the adopted version of a code, values of $\Upsilon$ may significantly disagree among calculations by different authors as may be noted from Figs. 12--18 of Aggarwal and Keenan \cite{oiv} for transitions of O IV. Some of these discrepancies arise due to  the non-convergence of either $\Omega$ (because of the limited number of partial waves included) or the integral in Eq. (5), because of the limited range of energy included, as clearly demonstrated in Figs. 2, 3 and 7 of  Aggarwal and Keenan \cite{nixi} for transitions in Ni XI. Resonances close to the thresholds can easily move left or right of a threshold due to differences in energy levels, and their positions sometimes considerably affect the calculations of $\Upsilon$, particularly at very low temperatures ($\sim$ 10$^3$--10$^4$ K). This is the main reason for the differences in the behaviour of the variation of $\Upsilon$ with T$_e$ as noted above for transitions of O IV. 

\section{SOURCES OF DATA}

Atomic data in the literature are often spread over a wide range of journals and this makes the task of even identifying data difficult for a user, apart from the problems of the assessment of the results as discussed above in sections 6 and 7. Additionally, due to a significant increase in computing and storage power during the past decade or so, a typical calculation generates so much data that no journal can publish these in their entirely, although some do provide a significant amount in their electronic versions. To overcome this difficulty, a few websites store a significant amount of data, for a variety of parameters and for a large range of ions. The most common and widely used websites, especially for collisional data, are: (i) CHIANTI: {\tt
http://www.damtp.cam.ac.uk/user/astro/chianti/}, (ii) ADAS: {\tt http://open.adas.ac.uk/},  (iii) APAP:  {\tt http://amdpp.phys.strath.ac.uk/} and (iv) CFADC: {\tt http://www-cfadc.phy.ornl.gov/}. Apart from the numerical atomic data, these websites also provide a wide range of computer programs for the applications of the data, and hence are extensively employed by those who generate, assess, and/or apply the atomic data to the modelling of plasmas. Therefore, particularly for a user who is not an expert in atomic data, these websites  and repositories of data are very helpful. However, websites also have some disadvantages. Data may be incomplete, are not assessed by independent experts, and are often unpublished. The other problem is that revising/updating the website is a continuous process, and since this is staff intensive, there can be a considerable delay before the (more accurate) published data are incorporated into the data repository. Furthermore, in rare cases data pertaining to the same calculation may be different on different websites, as recently highlighted by Aggarwal and Keenan \cite{fe25} for transitions of Fe XXV. As a consequence, the data  on a website may not always be the best available in the literature. Therefore, for an active researcher who wants to use the latest and/or the best available atomic data, there is no choice except to search for the data himself.

\section{ADVICE}

In this paper we have discussed a few examples (although there are many more) of  discrepancies for different atomic parameters. There can be several reasons for such discrepancies among different calculations. However, we would like to state that a calculation can only be considered to be the `best' available until a better one can be performed, as there is always scope for improvement.  For example, by (i) considering a larger model of an ion, (ii) including more CI in the generation of wavefunctions, (iii) widening the range of partial waves and/or energy, (iv) resolving resonances in a narrower energy mesh, (v) including radiation damping, (vi) including two-body relativistic operators (if not already done), and (vii) including the effects of higher lying ionization channels through pseudostates, which are particularly important for lighter H-like ions, such as He II and Li III. In conclusion, improving a calculation is always a continuous process and depends on the availability of the computer (and staff) resources, but large discrepancies (errors), particularly for  radiative rates and collisional data, can be avoided (or considerably reduced) by a few simple precautions as stated below for the benefit of the producers, assessors, and users of atomic data. 

\subsection{Producers}
\begin{enumerate}
\item Make as many comparisons as possible between different calculations for a variety of transitions, such as: allowed, forbidden, semi-forbidden, weak, and strong.
\item In the case of large discrepancies, try to understand and explain these without making assumptions.
\item Report results for collision strengths ($\Omega$), at least for a few transitions and at a few energies, so that a relationship with the subsequent results for  $\Upsilon$ can be formed. Most of the discrepancies in values of $\Upsilon$ can only be understood by knowing the corresponding differences in $\Omega$.
\end{enumerate}
\subsection{Assessors}
\begin{enumerate}
\item This is a difficult task to perform (although none is trivial), especially when it is not easy to assess even one's own work.
\item Assess what methods and assumptions have been used in a calculation.
\item Follow some basic guidelines, such as: behaviour of a transition, adequacy of the J/L and E ranges, inclusion (exclusion) of resonances, relativistic effects, etc.
\end{enumerate}
\subsection{Users}
\begin{enumerate}
\item The best situation is when only one set of data is available. However, this is often not the case.
\item If two or more data sets are available, and the authors do not fully and convincingly justify the improvements made, then use both (or more) sets of data and make your own
assessment. However, remember that the latest available calculation may not always be the best.
\item In case of doubt and/or suspicion, contact the authors.
\end{enumerate}

\section*{Acknowledgment}  KMA is grateful to  AWE Aldermaston for financial support and thanks the organisers of the IAEA-NFRI Technical Meeting for providing the hospitality.



\begin{thebibliography}{99}

\bibitem{fe25}  K.M.  AGGARWAL  and  F.P. KEENAN,   {\em Phys. Scr.}, {\bf 87}, pp (2013) in press
\bibitem {arxvii}  K.M. AGGARWAL  and F.P. KEENAN,   {\em Astron.  Astrophys.},  {\bf 441},  831 (2005)
\bibitem{ovii}   K.M. AGGARWAL  and F.P. KEENAN,  {\em Astron. Astrophys.} , {\bf 489}, 1377 (2008)
\bibitem {civ3}  A. HIBBERT, {\em Comput. Phys. Commun.},  {\bf 9},  141 (1975)
\bibitem{grasp0}  I.P. GRANT , B.J. McKENZIE, P.H.  NORRINGTON,  D.F. MAYERS  and N.C. PYPER,  {\em Comput. Phys. Commun.},  {\bf 21}, 207 (1980) 
\bibitem {ss}  W. EISSNER, M. JONES, H. NUSSBAUMER, {\em Comput. Phys. Commun.}, {\bf 8}, 271  (1974)
\bibitem {as}  N.R. BADNELL, {\em J. Phys. B},  {\bf 19},   3827 (1986)
\bibitem {mchf}  C. FROESE-FISCHER, G. TACHIEV, G. GAIGALAS, M.R. GODFROID, {\em Comput. Phys.  Commun. }, {\bf 176},  559 (2007)
\bibitem {mbpt}  M.S. SAFRONOVA, W.R. JOHNSON, U.I. SAFRONOVA, {\em Phys. Rev A},  {\bf 54}, 2850  (1996)
\bibitem {hfr}  R.D. COWAN, the Theory of Atomic Structure and Spectra, Univ. of California Press, 1981
\bibitem{fac}	M.F. GU,  {\em Can. J. Phys.},  {\bf 86}, 675   (2008)
\bibitem {hullac}  A. BAR-SHALOM, M. KLAPISCH, J. OREG,  {\em J. Quant. Spectrosc. Rad. Transfer},  {\bf 71}, 169  (2001)
\bibitem {rm}  K.A. BERRINGTON, W.B. EISSNER, P.H. NORRINGTON,  {\em Comput. Phys.  Commun.},  {\bf 92},   290 (1995)
\bibitem {dw}  W. EISSNER, M.J. SEATON, {\em J. Phys. B},  {\bf 5}, 2178  (1972)
\bibitem {c-like}  K.M. AGGARWAL, {\em Astrophys. J. Supl.},  {\bf 118}, 589  (1998)
\bibitem {ni-ions}  K.M. AGGARWAL, F.P. KEENAN, A.Z. MSEZANE, {\em At. Data Nucl. Data Tables},  {\bf 85},  453  (2003)
\bibitem{bbs}     C.P. BALLANCE, N.R. BADNELL,  E.S. SMYTH, {\em J. Phys.},  {\bf B 36},  3707 ( 2003)
\bibitem{oiii}   K. M. AGGARWAL and A. HIBBERT,  {\em J. Phys. B}, {\bf 24}, 3445 (1991)
\bibitem{kr35}    K.M. AGGARWAL  and  F.P. KEENAN,  {\em Phys. Scr.} , {\bf 86},  035302 (2009)
\bibitem {mg-like} K.M. AGGARWAL, V. TAYAL, G.P. GUPTA, F.P. KEENAN, {\em At. Data Nucl. Data Tables }, {\bf 93},  615  (2007)
\bibitem {hydro}  K. HAMADA, K.M. AGGARWAL, K. AKITA, A. IGARASHI, F.P. KEENAN, S. NAKAZAKI, {\em At. Data Nucl. Data Tables}, {\bf 96}, 481 (2010)
\bibitem{caxix}  K.M. AGGARWAL  and  F.P. KEENAN,   {\em Phys. Scr.},  {\bf 85},  025306 (2012)
\bibitem{nev}   K.M. AGGARWAL, {\em J. Phys. B},  {\bf 16},   2405 (1983)
\bibitem{mo34}   K.M. AGGARWAL, F.P. KEENAN, {\em Phys. Scr.},  {\bf 71}, 251  (2005)
\bibitem{gd37}    K.M. AGGARWAL, F.P. KEENAN, R. KISIELIUS, P.H. NORRINGTON, R.E. KING, G.J. PERT, S.J. ROSE, {\em Phys. Scr. }, {\bf 71},  356 (2005)
\bibitem{dpz}   F. DELAHAYE, A.K.   PRADHAN   and C.J.  ZEIPPEN,  {\em J. Phys.},  {\bf B39},   3465 (2006)
\bibitem{icft}   A.D. WHITEFORD, N.R. BADNELL,  C.P. BALLANCE, M.G. O'MULLANE, H.P. SUMMERS and A.L. THOMAS,  {\em J. Phys. B} {\bf 34}, 3179 (2001)
\bibitem{w47}   C.P. BALLANCE, D.C. GRIFFIN, {\em J. Phys. B},  {\bf 39},  3617  (2006)
\bibitem{damp}  D.C. GRIFFIN  and C.P. BALLANCE,  {\em J. Phys. B} ,{\bf  42},  235201 (2009)
\bibitem{tixxi}  K.M. AGGARWAL  and  F.P. KEENAN,  {\em Phys. Scr.},  {\bf 85},  065301 (2012)
\bibitem{lia}   K.M. AGGARWAL  and F.P. KEENAN,  {\em  At. Data Nucl. Data Tables},   (2012a) -- in press
\bibitem{lib}   K.M. AGGARWAL and F.P. KEENAN,  {\em  At. Data Nucl. Data Tables},  (2012b)  -- in press
\bibitem{ni19a}   K.M. AGGARWAL, F.P. KEENAN, {\em Pramana - J. Phys.},   {\bf 67},  C553 (2006)
\bibitem{ni19b}   K.M. AGGARWAL, F.P. KEENAN, {\em Pramana - J. Phys. },  {\bf 69},  209 (2007)
\bibitem{nixi}   K.M. AGGARWAL, F.P. KEENAN, {\em Eur. Phys. J. D},  {\bf 46},   205 (2008)
\bibitem{o3}    E. PALAY, S.N. NAHAR, A.K. PRADHAN, W. EISSNER, {\em Month. Not. R. Astron. Soc.}, {\bf 423}, L35 (2012) 
\bibitem{fe17} G.Y. LIANG, N.R. BADNELL, {\em Astron. Astrophys.} , {\bf 518}, A64 (2010)
\bibitem{fe14} G.Y. LIANG, N.R. BADNELL, J.R.C. LOPEZ-URRUTTIA, T.M. BAUMANN, G. DEL ZANNA, P.J. STOREY, H. TAWARA, J. ULRICH, {\em Astrophys. J. Supl.},  {\bf 190}, 322  (2010)
\bibitem{fe13}  P.J. STOREY, C.J. ZEIPPEN, {\em Astron. Astrophys.} , {\bf 511}, A78 (2010)
\bibitem{fe11}  G. DEL ZANNA, P.J. STOREY, H.E. MASON, {\em Astron. Astrophys.} , {\bf 514}, A40 (2010)
\bibitem {kr32}  K.M. AGGARWAL, F.P. KEENAN, K. D. LAWSON, {\em At. Data Nucl. Data Tables},  {\bf 95},  607 (2009)
\bibitem{ni19c}   K.M. AGGARWAL, F.P. KEENAN, {\em Astron. Astrophys.},  {\bf 460},  959  (2006)
\bibitem{ni19d}   A. HIBBERT, M. Le DOURNEUF, M. MOHAN, {\em At. Data Nucl. Data Tables }, {\bf 53},  23 (1993)
\bibitem{ni17a}   R. DAS, N.C. DEB, K. ROY, A.Z. MSEZANE, {\em Phys. Scr.},  {\bf 67},   401 (2003)
\bibitem{fe15}   K.M. AGGARWAL, V. TAYAL, G.P. GUPTA and F.P. KEENAN,  {\em  At. Data Nucl. Data Tables},  {\bf 93},  615 (2007)
\bibitem {al-like}  G.P. GUPTA, K.M. AGGARWAL,  A.Z. MSEZANE, {\em Physical  Review},  {\bf A70},   036501 (2004)
\bibitem{saf}   U.I. SAFRONOVA, C. NAMBA, J.R. ALBRITTON, W.R. JOHNSON, M.S. SAFRONOVA, {\em Physical  Review}, {\bf A65} , 022507 (2002)
\bibitem {kr-ions}  K.M. AGGARWAL, F.P. KEENAN, K. D. LAWSON, {\em At. Data Nucl. Data Tables},  {\bf 94},   323 (2008)
\bibitem{fe16a}   K.M. AGGARWAL, F.P. KEENAN, {\em Astron. Astrophys.},  {\bf 463},  399 (2007)
\bibitem{fe9b}   N. VERMA, A.K.S. JHA, M. MOHAN, {\em Astrophys. J. Suppl.}, {\bf 164},  297 (2006)
\bibitem{fe9a}   K.M. AGGARWAL, F.P. KEENAN, T. KATO, I. MURAKAMI, {\em Astron. Astrophys.},  {\bf 460},   331 (2006)
\bibitem{fe16b}   K.M. AGGARWAL, F.P. KEENAN, {\em Astron. Astrophys.},  {\bf 450},   1249 (2006)
\bibitem{fe16c}   M. CORNILLE, J. DUBAU, H.E. MASON, C. BLANCARD, W.A. BROWN, {\em Astron. Astrophys.}, {\bf 320},  333  (1997)
\bibitem{fe16d}  W. EISSNER, M.E. GALAVIS, C. MENDOZA, C.J. ZEIPPEN, {\em Astron. Astrophys. Suppl.},  {\bf 136},  385 (1999)
\bibitem{fe16e}   M.A. BAUTISTA, {\em J. Phys. B}, {\bf 33},   71 (2000)
\bibitem{fe16f}  K.M. AGGARWAL, F.P. KEENAN, {\em J. Phys. B},  {\bf 41},   015701 (2008)
\bibitem{mgxi}  K.M. AGGARWAL  and  F.P. KEENAN,  {\em Phys. Scr.},  {\bf 85},  025305 (2012)
\bibitem{lb}      G.Y. LIANG and N.R. BADNELL,  {\em Astron. Astrophys.},  {\bf 528},  A69 (2011)
\bibitem{ar16} A.D. WHITEFORD,  N.R. BADNELL,  C.P. BALLANCE,  S.D. LOCH,  M.G. O'MULLANE  and H.P. SUMMERS,  {\em J. Phys. B},  {\bf 35},  3729 (2002)
\bibitem{oiv}   K.M. AGGARWAL, F.P. KEENAN, {\em Astron. Astrophys.},  {\bf 486},  1053 (2008)


\end{thebibliography}
\end{document}